\documentclass[journal]{IEEEtran}
\usepackage{multirow}
\usepackage{graphicx}
\usepackage{amsmath}
\usepackage{color}
\usepackage{cite}

\newcommand\boldblue[1]{\textcolor{blue}{\textbf{#1}}}

\begin{document}

\title{Image Processing Operations Identification via Convolutional Neural Network}

\author{Bolin Chen,~\IEEEmembership{}
	  Haodong Li,~\IEEEmembership{Student Member,~IEEE,}
        Weiqi Luo*,~\IEEEmembership{Senior Member,~IEEE}

\thanks{B. Chen and W. Luo (corresponding author) are with the School of Data and Computer Science, and Guangdong Key Laboratory of Information Security Technology, Sun Yat-sen University, Guangzhou 510006, P.R. China (e-mail: chenbl8@mail2.sysu.edu.cn, luoweiqi@mail.sysu.edu.cn).}
\thanks{H. Li is with the College of Information Engineering, Shenzhen University, Shenzhen 518052, P.R. China.}
}

\markboth{}{}

\maketitle

%:
\begin{abstract}

In recent years, image forensics has attracted more and more attention, and many forensic methods have been proposed for identifying image processing operations. Up to now, most existing methods are based on hand crafted features, and just one specific operation is considered in their methods. In many forensic scenarios, however,  multiple classification for various image processing operations is more practical. Besides, it is difficult to obtain effective features by hand for some image processing operations. In this paper, therefore, we propose a new convolutional neural network (CNN) based method to adaptively learn discriminative features for identifying typical image processing operations. We carefully design the high pass filter bank to get the image residuals of the input image, the channel expansion layer to mix up the resulting residuals, the pooling layers, and the activation functions employed in our method. The extensive results show that the proposed method  can outperform the currently best method based on hand crafted features and three related methods based on CNN for image steganalysis and/or forensics, achieving the state-of-the-art results. Furthermore, we provide more supplementary results to show the rationality and robustness of the proposed model.

\end{abstract}


\IEEEpeerreviewmaketitle

\section{Introduction}
\label{Sec:Introduction}

\IEEEPARstart{W}{ith} the rapid development of image processing technology, it is much easier to modify digital images without leaving any perceptible artifacts than ever before. Nowadays, the abuse of tampered images would lead to many potential serious moral, ethical and legal problems. Therefore, image forensics \cite{StammWL2013}  has attracted increasing attention.

Usually, any image processing operation would inevitably modify and distort some inherent statistics within  original images. Therefore, it is possible to detect resulting images via analyzing the artifacts left by the image operation. Up to now, most existing methods just consider only one specific image processing operation. For example, identifying JPEG compression history \cite{fan2003identification,farid2009exposing,luo2010jpeg,bianchi2012detection}, contrast enhancement \cite{stamm2008blind,cao2014contrast}, resampling \cite{popescu2005exposing,mahdian2008blind,li2013moment}, median filtering \cite{kirchner2010detection,yuan2011blind,kang2012robust,chen2013blind}, image splicing \cite{shi2007steganalysis,he2012digital,zhao2015passive}, etc. Such methods usually analyze the specific artifacts left by the targeted operation,  and then design features according to the artifacts, and finally employ binary classification for detection. Although promising performance can be achieved for some targeted operations, such methods usually suffer from poor performance for detecting other operations. For example, a powerful method for detecting JPEG compression may not be suitable for detecting median filtering or Gamma correction. Instead of targeted methods, universal method that can identify various image processing operations is worth the effort. Several related works have been proposed until now. For instance, by modeling image processing operations as steganography, Qiu et al. \cite{qiu2014universal} proposed to use steganalytic features, such as SPAM \cite{pevny2010steganalysis}, SRM \cite{fridrich2012rich} and LBP \cite{shi2012textural}, to identify six typical image processing operations. Fan et al. \cite{fan2015general} adopted Gaussian mixture models to model the statistics of images processed by different image operations. Recently, Li et al. \cite{li2016identification} proposed a compact universal feature set from SRM \cite{fridrich2012rich} to identify 11 typical image processing operations.

The above methods are all based on hand crafted features, meaning that a lot of efforts are required for constructing effective features. Another novel solution is to use the convolutional neural network (CNN) \cite{lecun1998gradient} to learn discriminative features from the data. It is well known that CNN has achieved great success in various research fields, especially in computer vision \cite{KrizhSH2012,simonyan2014very}.
Up to now, several CNN based methods have been proposed for image steganalysis and forensics.
For instance, Qian et al. \cite{qian2015deep} proposed a Gaussian-Neuron Convolutional Neural Network and achieved comparable performance with SRM for image steganalysis  \cite{fridrich2012rich}. Xu et al. \cite{xu2016structural} proposed a CNN equipped with batch normalization \cite{ioffe2015batch} for image steganalysis, and used this network in \cite{xu2016structural} as base learners and combined ensemble strategies to improve the performance \cite{xu2016ensemble}. Recently, Ye et al. \cite{ni2017deep} proposed a CNN that adopted truncated linear unit as activation function and incorporated the knowledge of selection channel to achieve better performance compared with several typical steganographic algorithms.
In image forensics, Chen et al.\cite{chen2015median} proposed a CNN model to detect those images median filtering. Bayar et al. \cite{bayar2016deep} developed a new convolutional layer to suppress the image contents for detecting some typical image processing operations,  Cozzolino et al. \cite{cozzolino2017recasting} showed that a class of residual-based descriptors in forensics can be regarded as a simple constrained CNN. Motivated by many breakthrough results in computer vision, the focus of image steganalysis and forensics is gradually shifting on deep learning.

In this paper, we carefully design a new CNN based method for detecting various typical image processing operations. In the proposed method, we first convert the input image into residuals to suppress the influence of image contents, and then use a convolutional layer to increase the channel number. After then, we employ six similar layer groups  to obtain the high level features of the input image. Finally, we feed the resulting features into the full connect layer for classification. Extensive experiments have shown that the proposed method can achieve the state-of-the-art results compared with currently the best method based on hand crafted features \cite{li2016identification} and three other related CNN based methods \cite{chen2015median,bayar2016deep,xu2016structural} for image steganalysis and/or forensics. Moreover, we have validated the rationality and  robustness of the proposed model with more supplementary results.

The rest of the paper is organized as follows. Section \ref{Sec:Proposed} describes the structure of the proposed network. Section \ref{Sec:Experiment} shows the experimental results and discussions. Finally, the concluding remarks are given in Section \ref{Sec:Conclusion}.

\section{The Proposed Method}
\label{Sec:Proposed}
%In this section, we will describe the overall architecture of the proposed CNN based model, and discuss some details about the model.

\begin{figure*}
  \centering
    \includegraphics[width=17.2cm]{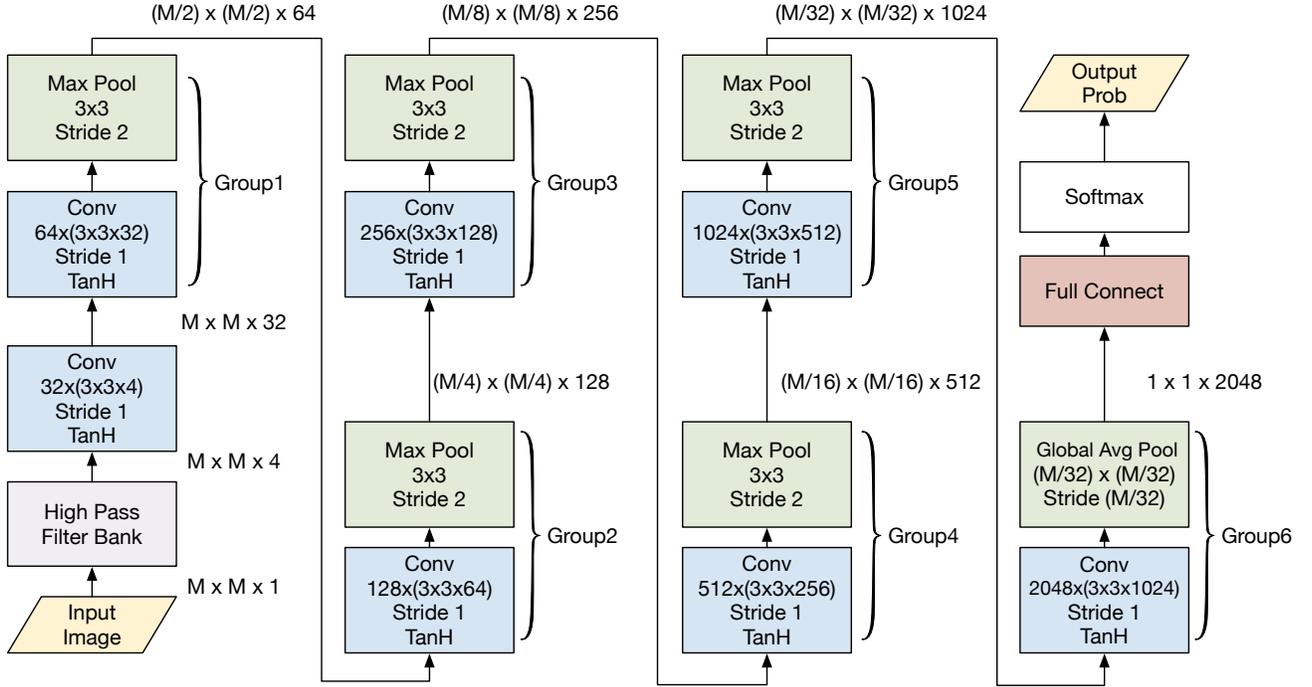}

   \caption{The architecture of the proposed model.}
   \label{fig:overall_architecture}
\end{figure*}

\begin{figure}
  \centering
    \includegraphics[width=8cm]{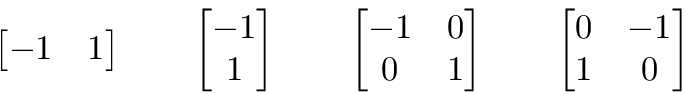}

   \caption{Four high pass filters used in the proposed model.}
   \label{fig:filter_bank}
\end{figure}

The architecture of the proposed model is illustrated in Fig. \ref{fig:overall_architecture}.
Assume that the model input is a gray-scale image with the size of $M\times M$ ($M$ a multiple of 32). Based on our previous studies  \cite{li2016identification}, the artifacts introduced by various image processing operations are easier to capture in the image residual domain. Thus, the proposed model firstly transforms the input image into residuals with some high pass filters. Up to now, many high pass filters can be borrowed from the SRM \cite{fridrich2012rich} that is widely used in image steganalysis and forensics. Based on our extensive experiments, we found that the high pass filters with the first order usually have better detection performance than those with the second or higher orders in the proposed model. Furthermore, we also found that the detection performance would not increase significantly with increasing the number of the employed filters. Considering the trade off between the  model complexity and the detection performance, therefore, we just use the four simplest filters to capture the adjacent pixel differences in four different directions as shown in Fig \ref{fig:filter_bank}, and achieve very good detection results.

And then we use an ``channel expansion layer'' to process the resulting residuals and  increase the channel number of feature maps from 4 to 32 in this layer. The resulting feature maps are then fed into six similar and typical layer groups to obtain high level features for identifying the images after various image processing operations. In each of the six groups, it contains a convolutional layer and a pooling layer. The convolutional layer aims to double the channel number, while the pooling layer (except for the last one) tries to downsample the feature maps by 2 along the image width and height simultaneously. Please note that the pooling layer in the last group is different, it downsamples the feature maps to 1 along image width and height via the average pooling. This type of pooling layer is known as ``global average pooling'', which is introduced some typical deep learning based works such as \cite{lin2013network}. Compared with other commonly used methods, such as max pooling or average pooling, the global average pooling can improve performance.

In the proposed model, all the convolutional layers are equipped with the activation function TanH, which is more suitable for the proposed network compared with other popular activation functions such as ReLu and Sigmoid. Besides, all the convolutional layers and pooling layers use the smallest size kernel with a center (i.e.,$3\times3$) except for the global average pooling, since small kernel size have fewer parameters which  helps train the network faster and prevent overfitting. Besides, based on previous study \cite{simonyan2014very}, the size of $3\times3$ is the smallest one to capture the notion of left/right, up/down and center, and the use of such small convolution filters in all layers usually has good results for image recognition.

Please note that some comparative experiments and discussions about above setups will be given in Section \ref{Subsec:comparison_structure}.

\section{Experimental Results}
\label{Sec:Experiment}
In this section, we first describe the image dataset used in our experiments and the detail setups of the proposed model, and then present extensive experiments to show the effectiveness of the proposed model.

\subsection{Image Dataset}

\begin{table*}[!t]

\renewcommand{\arraystretch}{1.0}

\caption{The parameters of image processing operations used in the experiments}
\label{table_parameters}
\centering
\begin{tabular}{p{4cm}|p{8cm}}
\hline
Operation type & Parameters\\
\hline
Gamma correction(GC) & $\gamma$: 0.5, 0.6, 0.7, 0.8, 0.9, 1.2, 1.4, 1.6, 1.8, 2.0 \\
\hline
Histogram equalization(HE) & n/a \\
\hline
Unsharp masking sharpening(UM) & $\sigma$: $0.5 - 1.5$; $\lambda$: $0.5 - 1.5$ \\
\hline
Mean filtering(MeanF) & window size: $3\times3$, $5\times5$, $7\times7$ \\
\hline
Gaussian filtering(GF) & window size: $3\times3$, $5\times5$, $7\times7$; $\sigma$: 0.8 - 1.6 \\
\hline
Median filtering(MedF) & window size: $3\times3$, $5\times5$, $7\times7$ \\
\hline
Wiener filtering(WF) & window size: $3\times3$, $5\times5$, $7\times7$ \\
\hline
\multirow{2}{*}{Scale(Sca)} & up-sampling: 1, 3, 5, 10, 20, 30, 40, 50, 60, 70, 80, 90 (\%) \\
 & down-sampling: 1, 3, 5, 10, 15, 20, 25, 30, 35, 40, 45 (\%) \\
\hline
Rotation(Rot) & degree: 1, 3, 5, 10, 15, 20, 25, 30, 35, 40, 45 ($^\circ$) \\
\hline
JPEG & quality factor: $75 - 99$ \\
\hline
JPEG 2000(JP2) & compression ratio: $2.0 - 8.0$ \\
\hline
\end{tabular}
\end{table*}

In our experiments, we firstly collect 40,000 images in raw format from different cameras, and transform them into gray-scale with the size of $512\times512$ as original images. Eleven typical image processing operations are considered, including Gamma correction (GC), Histogram equalization (HE), Unsharp masking sharpening (UM), Mean filtering (MeanF), Gaussian filtering (GF), Median filtering (MedF), Wiener filter (WF), Scaling (Sca), Rotation (Rot), JPEG and JPEG 2000 (JP2). As it did in \cite{li2016identification}, these operations are performed on each original image with a random parameter selected in Table \ref{table_parameters}.
In all, therefore, we obtain 12 classes (including original one) of images, each of which contains   40,000 images. In each experiment, we firstly  divide the original images into training data set (26,000 images), validation data set (4,000 images) and testing data set (10,000 images), and then select the corresponding images after some image processing operations as corresponding data set for binary and multiple classification. In our experiments, we randomly divide the training, validation and testing data three times and report the average results in the following sections.

\subsection{Implementation Details} % ¸Ä

We implemented the proposed CNN based model using tensorflow \cite{abadi2016tensorflow}. Instead of the typical vanilla SGD, we employed Nesterov Momentum \cite{sutskever2013importance} to train our network since it learnt faster and performed better based on our experiments. In the experiments, the momentum was set as 0.9, and  L2 regularization was used. The corresponding weight decay was 0.0005. We initialized all the weights using Gaussian distribution with 0 mean and standard deviation 0.01, and initialized all the biases with 0. In the training stage, the batch size was 64. We shuffled the training set between epochs.

Besides, step decay of learning rate was used in our method. We gradually reduced the learning rate during training. We divided the learning rate by 10 when the validation accuracy stopped improving, and we stop the training after reducing the learning rate three times. For the proposed network, the initialization of learning rate is 0.01. In our experiments, it decreased at iterations 60,000, 100,000 and finally stopped at iterations 120,000. Please note that the schedule configuration of learning rate is different for different networks. To achieve good performance, we used similar rules as mentioned above to reduce the learning rate for other networks in the experiments. Some detailed settings are not given here.

\subsection{Comparison with Existing Works}
\label{sub:compared_others}

In this subsection, we compare the proposed method with some related works, including the current best method based on handcrafted features \cite{li2016identification}, and three other CNN based methods including Chen's method \cite{chen2015median}, Bayar's method \cite{bayar2016deep} and Xu's method \cite{xu2016structural}. Please note that Xu's method \cite{xu2016structural} is originally designed for image steganalysis. However, based on our previous analysis  \cite{qiu2014universal}, we found that some modern steganlytic features/models are very effective for identifying some image forensics applications. Recently, the work \cite{zhan2017image} employs the transfer learning and the CNN architecture in \cite{xu2016structural} for image processing operations classification, and show that both methods \cite{zhan2017image} and \cite{xu2016structural} have good and similar detection results. In this paper, we also include Xu's method \cite{xu2016structural} for comparative study.

For a fair comparison, the size of input image should be the same for all methods. Thus we have to slightly modify the corresponding input layer and the layer that aggregates the output feature maps of convolutional layers  due to the image size and/or the memory limitation, while preserve other layers as they are for other CNN based methods. In addition, we improve the Bayar's method \cite{bayar2016deep} via just enforcing the constraint once the CNN weights are initialized other than enforcing the constraint in each iteration, and obtain much better detection results based on our experiments. In the following, binary classification and multi-class classification have been evaluated respectively.

\subsubsection{Binary Classification}
\label{subsubsec:binary}

\begin{table*}[!t]
\renewcommand{\arraystretch}{1.0}
% if using array.sty, it might be a good idea to tweak the value of
% \extrarowheight as needed to properly center the text within the cells
\caption{Detection accuracy (\%) for different image processing operations. The best accuracy for each operation is highlighted and labeled with an asterisk (*).}
\label{binary_classification}
\centering
\begin{tabular}{l|cccccccccccc}
\hline
 & GC &HE & UM & MeanF & GF & MedF & WF & Sca & Rot & JPEG & JP2 & Average\\
\hline
The proposed method & \boldblue{96.40*} &  99.88 & \boldblue{99.17*} & 99.94  & 99.96 & 99.96 & 99.92 & \boldblue{97.39*} & \boldblue{99.19*} & \boldblue{99.72*} & 99.75 & \boldblue{99.21*}\\
Li's Method \cite{li2016identification} & 94.28 & \boldblue{99.90*} & 97.06 & \boldblue{99.97*} & \boldblue{99.98*} & \boldblue{99.99*} & \boldblue{99.96*} & 96.21 & 99.04 & 98.79 & \boldblue{99.82*}  & 98.64\\
Chen's Method \cite{chen2015median} & 79.16 & 98.48 & 96.52 & 99.85 & 99.84 & 99.83 & 99.89 & 93.37 & 98.26 & 98.54 & 99.07  & 96.62\\
Bayar's Method \cite{bayar2016deep} & 64.48 & 98.15 & 91.71 & 99.83 & 99.86 & 99.23 & 99.69 & 93.99 & 97.34 & 97.96 & 98.26 & 94.59\\
Xu's Method \cite{xu2016structural} & 76.03 & 97.61 & 94.87 & 99.92 & 99.89 & 99.89 & 99.77 & 95.11 & 98.59 & 98.97 & 99.68 & 96.39\\
\hline

\end{tabular}
\end{table*}

In this experiment, we try to identify whether a given image is original or modified by a certain image processing operation. The average detection results are shown in Table \ref{binary_classification}. From Table \ref{binary_classification},  we can observe that all the methods can obtain satisfactory results (larger than 91\%) for all image processing operations except for Gamma correction. Overall, the proposed CNN based method usually works better than the existing CNN based works (i.e. \cite{chen2015median}, \cite{bayar2016deep} and \cite{xu2016structural}), especially for identifying the operations of GC, UM and Sca. For instance, we obtain 96.4\% detection accuracy for the GC, while less than 80\%  for the other CNN based works.
For the current best work \cite{li2016identification}, we can obtain similar results for all image processing operations. On average, the proposed CNN based method outperforms the work \cite{li2016identification} slightly for the binary classification (around 0.5\% improvement), please refer to the final column of  Table \ref{binary_classification}.

\subsubsection{Multiple Classification}

\begin{table*}[!t]

\renewcommand{\arraystretch}{1.0}

\caption{Confusion matrix for identifying  image processing operations using the proposed method. The asterisk ``*'' here means that the corresponding value is less than 1\%. }
\label{cm_proposed_method}
\centering
\begin{tabular}{c|cccccccccccc}
\hline
Actual/Predicted & Orig & GC &HE & UM & MeanF & GF & MedF & WF & Sca & Rot & JPEG & JP2 \\
\hline
Orig & \textbf{97.48} & 1.36 & * & *  & * &  * & * & * & * &  * & * & * \\
GC & 3.99 &  \textbf{95.13} & * & *  & * &  * & * & * & * &  * & * & * \\
HE & * &  * & \textbf{99.33} & *  & * &  * & * & * & * &  * & * & * \\
UM & * &  *& * & \textbf{98.57}  & * &  * & * & * & * &  * & * & * \\
MeanF & * &  * & * & *  & \textbf{99.12} & * & * & * & * &  * & * & * \\
GF & * &  * & * & *  & * &  \textbf{99.36} & * & * & * &  * & * & * \\
MedF & * &  * & * & *  & * &  * & \textbf{99.80} & * & * &  * & * & * \\
WF & * &  * & * & *  & * &  * & * & \textbf{99.31} & * &  * & * & * \\
Sca & 1.45 &  * & * & *  & * &  * & * & * & \textbf{98.30} &  * & * & * \\
Rot & * &  * & * & *  & * &  * & * & * & * &  \textbf{99.85} & * & * \\
JPEG & * &  * & * & *  & * &  * & * & * & * &  * & \textbf{99.18} & * \\
JP2 & * &  * & * & *  & * &  * & * & * & * &  * & * & \textbf{99.53} \\
\hline

\end{tabular}
\end{table*}

\begin{table*}[!t]

\renewcommand{\arraystretch}{1.0}

\caption{ Confusion matrix for identifying  image processing operations using Li's Method\cite{li2016identification}. The asterisk ``*'' here means that the corresponding value is less than 1\%. }
\label{cm_method_1}
\centering
\begin{tabular}{c|cccccccccccc}
\hline
Actual/Predicted & Orig & GC &HE & UM & MeanF & GF & MedF & WF & Sca & Rot & JPEG & JP2 \\
\hline
Orig & \textbf{92.46} & 3.28 & * & 1.67  & * &  * & * & * & 1.66 &  * & * & * \\
GC & 7.45 &  \textbf{89.42} & 1.70 & *  & * &  * & * & * & * &  * & * & * \\
HE & * &  1.57 & \textbf{98.38} & *  & * &  * & * & * & * &  * & * & * \\
UM & 1.86 &  *& * & \textbf{97.14}  & * &  * & * & * & * &  * & * & * \\
MeanF & * &  * & * & *  & \textbf{96.59} & 2.40 & * & * & * &  * & * & * \\
GF & * &  * & * & *  & 1.80 & \textbf{98.02} & * & * & * &  * & * & * \\
MedF & * &  * & * & *  & * &  * & \textbf{99.80} & * & * &  * & * & * \\
WF & * &  * & * & *  & 1.15 &  * & * & \textbf{98.14} & * &  * & * & * \\
Sca & 4.87 &  * & * & *  & * &  * & * & * & \textbf{92.87} & 1.51 & * & * \\
Rot & 1.30 &  * & * & *  & * &  * & * & * & 2.65 & \textbf{95.77} & * & * \\
JPEG & 1.79 &  * & * & *  & * &  * & * & * & * &  * & \textbf{97.64} & * \\
JP2 & * &  * & * & *  & * &  * & * & * & * &  * & * & \textbf{99.50} \\
\hline

\end{tabular}
\end{table*}

\begin{table*}[!t]
\renewcommand{\arraystretch}{1.0}
\caption{ Confusion matrix for identifying  image processing operations using  Chen's Method \cite{chen2015median}. The asterisk ``*'' here means that the corresponding value is less than 1\%.}
\label{cm_method_2}
\centering
\begin{tabular}{c|cccccccccccc}
\hline
Actual/Predicted & Orig & GC &HE & UM & MeanF & GF & MedF & WF & Sca & Rot & JPEG & JP2 \\
\hline
Orig & \textbf{81.68} & 11.63 & * & 2.15  & * &  * & * & * & 2.11 & 1.25 & * & * \\
GC & 30.84 &  \textbf{58.99} & 3.21 & 3.82  & * &  * & * & * & 1.13 & 1.26 & * & * \\
HE & * &  2.72 & \textbf{95.62} & *  & * &  * & * & * & * &  * & * & * \\
UM & 2.80 & 3.77 & 1.28 & \textbf{91.76}  & * &  * & * & * & * &  * & * & * \\
MeanF & * &  * & * & *  & \textbf{97.16} & 2.48 & * & * & * &  * & * & * \\
GF & * &  * & * & *  & 2.85 &  \textbf{96.97} & * & * & * &  * & * & * \\
MedF & * &  * & * & *  & * &  * & \textbf{99.14} & * & * &  * & * & * \\
WF & * &  * & * & *  & 1.58 &  * & * & \textbf{97.37} & * &  * & * & * \\
Sca & 6.39 & 1.73 & * & *  & * &  * & * & * & \textbf{88.27} & 2.17 & * & * \\
Rot & * &  * & * & *  & * &  * & * & * & 1.27 &  \textbf{97.30} & * & * \\
JPEG & 1.23 & * & * & *  & * &  * & * & * & * &  * & \textbf{96.90} & * \\
JP2 & * & * & * & *  & * &  * & * & * & * &  * & * & \textbf{97.17} \\
\hline

\end{tabular}
\end{table*}

\begin{table*}[!t]
\renewcommand{\arraystretch}{1.0}

\caption{Average results (\%) along the diagonal line of the corresponding confusion matrix. The best accuracy among the five methods is highlighted and labeled with an asterisk (*).}
\label{diagonal_average}
\centering
\begin{tabular}{c|ccccc}
\hline
Method & The proposed Method & Li's Method\cite{li2016identification} & Chen's Method\cite{chen2015median} & Bayar's Method\cite{bayar2016deep} & Xu's Method\cite{xu2016structural} \\
\hline
Accuracy & \boldblue{98.75*} & 96.31 & 91.53 & 87.22 & 85.92  \\

\hline

\end{tabular}
\end{table*}

In many forensic scenes, multiple classification is more practical and more difficult compared with binary classification. In this experiment, we try to identify 11 typical operations shown in Table \ref{table_parameters}. The average confusion matrices for the proposed method, Li's method \cite{li2016identification} and Chen's method \cite{chen2015median} are shown in Table \ref{cm_proposed_method}, Table \ref{cm_method_1} and  Table \ref{cm_method_2} respectively.
From Table \ref{cm_proposed_method}, we can observe that the proposed method can effectively identify most image processing operations with a high detection accuracy. All values along the diagonal line of the confusion matrix are larger than 95\%, and almost all values located at non-diagonal line (i.e. false detection rates) are less than 1\%. The false detection rates are relatively larger for the GC and Sca, which is consistent with the results for binary classification shown in Table \ref{binary_classification}. From Table \ref{cm_method_1}, however, it is observed that the detection performance using the method \cite{li2016identification} will drop a lot. Taking the first column of the Table \ref{cm_method_1} for instance, more images after the GC, UM, Sca, Rot and JPEG are falsely predicted as original ones. The detection performance becomes even poorer with the Method \cite{chen2015median}, refer to Table \ref{cm_method_2} for details.
Due to the page limitation, we will not give the confusion matrix of Bayar's method \cite{bayar2016deep}  and Xu's method \cite{xu2016structural} here. Instead, we show the average results along the diagonal values of the corresponding confusion matrices for the five methods in Table \ref{diagonal_average}. From Table \ref{diagonal_average}, we can observe that the proposed CNN based method outperforms the other works, and improves the current best work \cite{li2016identification} over 2\%, which is a significant improvement on multiple classification about 11 image processing operations.

\subsection{Validation of the Designs of Model Architecture}
\label{Subsec:comparison_structure}

\begin{figure}
  \centering
    \includegraphics[width=9cm]{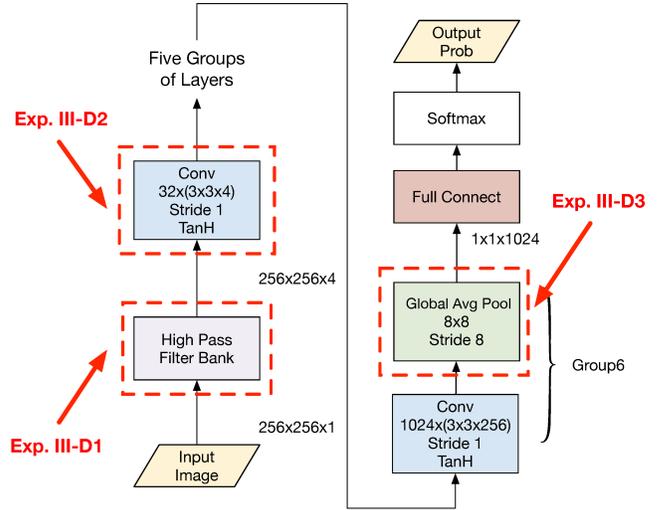}

   \caption{The network components related to each experiment in the Subsection \ref{Subsec:comparison_structure}.}
   \label{fig:exp_component}
\end{figure}

In this subsection, we try to present some experimental results to further validate the rationality of the proposed model. As illustrated in Fig. \ref{fig:exp_component}, three parts of the proposed model have been considered, including the high pass filter bank, the ``channel expansion layer'' and the last pooling layer. Besides, we also consider the activation functions used in the proposed structure. The corresponding results for multiple classification and analysis are shown in the four following subsections.

\subsubsection{About the High Pass Filter Bank}

\label{Subsubsec:high_pass}

\begin{figure}
  \centering
    \includegraphics[width=9cm]{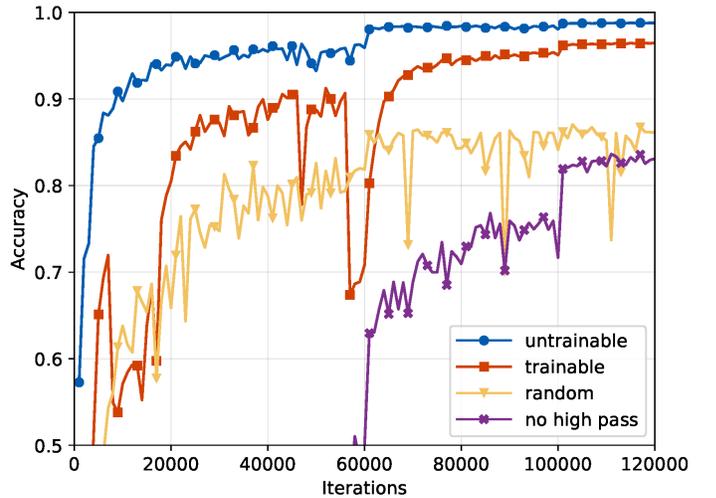}

   \caption{Comparison of different settings in the High Pass Filter Bank}
   \label{fig:filter_comparison}
\end{figure}

As described in Section \ref{Sec:Proposed}, we use four fixed high pass filters to get the image residuals of the input images. Some related works, such as \cite{ni2017deep} and \cite{rao2016deep}, usually use some high pass filters to initialize the first convolutional layer in their models, and make the weights of the filters trainable. In this experiment, therefore, we consider the four different cases about the High Pass Filter Bank, including using the fixed four filters (denoted as ``untrainable''), using the four filters for initialization (``trainable''), using the same size of filters with random initialization (``random'') and  removing the  high pass filter bank (``no high pass'', namely just copying the input image in the four channels).

The experimental results are shown in Fig. \ref{fig:filter_comparison}. From Fig.\ref{fig:filter_comparison},  we can observe that the proposed method (i.e. using fixed high pass filters) can achieve the best performance. Using the trainable high pass filter bank can achieve similar detection accuracy with ours when the number of iteration is large. However, the detection accuracy seems unstable during the early stage of the training. Replacing the high pass filter bank with random initialized convolutional layer shows poor performance (lower than 90\%). In addition, its performance seems unstable during the whole training stage. Removing the high pass filter has the lowest accuracy among all the approaches and it converges much slower than others.

\subsubsection{About the Channel Expansion Layer}
\label{Subsubsec:additional_conv}

\begin{figure}
  \centering
    \includegraphics[width=8cm]{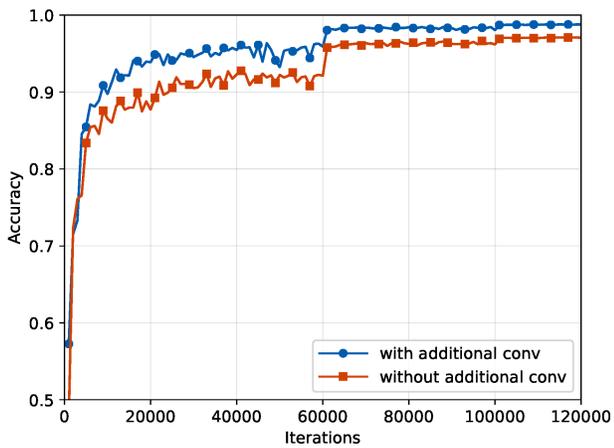}

   \caption{Comparison of the Channel Expansion layer.}
   \label{fig:additional_conv_comparison}
\end{figure}

The channel expansion layer, the convolutional layer right after the high pass filter bank, aims to mix up the residuals produced by the high pass filter bank and increase the channel number. We try to use this layer to combine the information of different residuals and  provide more input features for the six subsequent layer groups. In this experiment, we  show the detection performance if removing this layer. The experimental results are shown in Fig. \ref{fig:additional_conv_comparison}. From  Fig. \ref{fig:additional_conv_comparison}, we can observe that the channel expansion layer can slightly improve the detection performance. In addition, we try to add a pooling layer after the channel expansion layer, making the network has 7 similar groups of layers. Our experiments show that this  leads to a worse performance.

\subsubsection{About the Last Pooling Layer}
\label{Subsubsec:last_pooling}

\begin{figure}
  \centering
    \includegraphics[width=8cm]{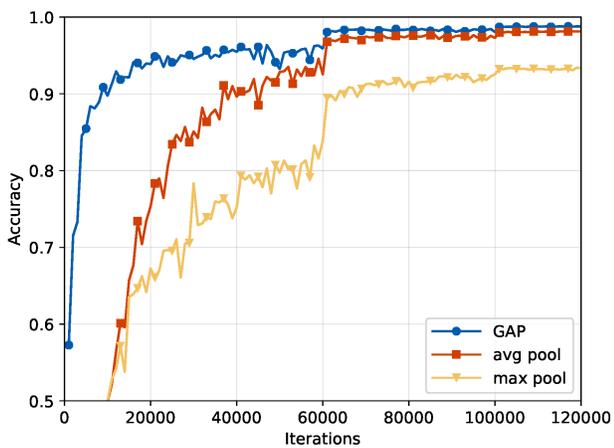}

   \caption{Comparison of different last pooling layer.}
   \label{fig:pooling_comparison}
\end{figure}

Pooling layer is a important component of CNN. Usually, CNNs adopt pooling with stride 2 to downsample input feature maps. Most CNN based methods in image forensics and steganalysis such as \cite{qian2015deep,chen2015median,bayar2016deep} simply use this type of pooling layer throughout the whole network. Only a few works such as \cite{xu2016structural} use global average pooling (GAP) in the last pooling layer. Different from conventional pooling layer, GAP aims to reduce the input feature maps to size of $1\times1$. In our method, we also adopt GAP in the last pooling layer.

In this experiment, we evaluate different pooling methods in the last pooling layer, including max pooling with stride 2, average pooling layer with stride 2, and GAP. The comparative results are shown in Fig. \ref{fig:pooling_comparison}. From Fig. \ref{fig:pooling_comparison}, the GAP works better than the two other methods, while the common max pooling has the worst performance.

Although average pooling outperform max pooling in the last pooling layer, it does not mean that using average pooling instead of max  pooling in the whole network will always has better performance. In fact, according to our experiments, replacing all the max pooling in the network with average pooling will make the training stuck from beginning.

\subsubsection{About the Activation Functions}
\label{Subsubsec:activation}

\begin{figure}
  \centering
    \includegraphics[width=8cm]{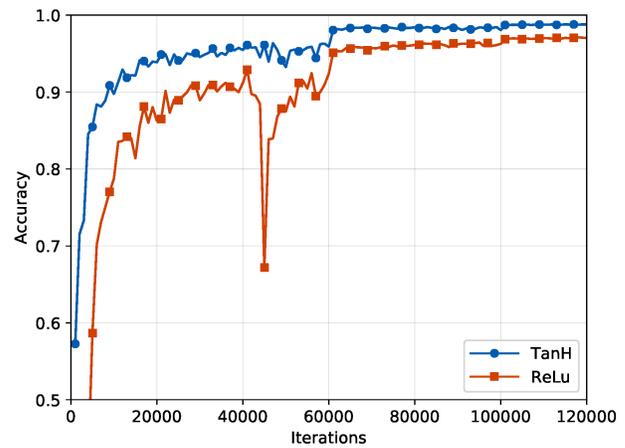}

   \caption{Comparison of Different Activation Functions.}
   \label{fig:activation_comparison}
\end{figure}

Activation function is another important issue in CNN, and the commonly used activation functions includes  Sigmoid, TanH and ReLu \cite{nair2010rectified}. In this experiment, we evaluate the proposed model with the three activation functions. The experimental results are shown in Fig. \ref{fig:activation_comparison}. Please note that we do not show the results of the Sigmoid function in this figure, since using the Sigmoid will stall the learning at the beginning, which means that the Sigmoid function is useless in the proposed model. From Fig. \ref{fig:activation_comparison}, we can observe that TanH  performs better and more stable than ReLu.

It is well known that ReLu usually produces sparser features compared with TanH. In many tasks in computer vision, the key features \cite{glorot2011deep} to distinguish different classes of objects tend to be prominent. Therefore, ReLu usually works better than TanH in these cases. However, the investigated forensic problem is quite different, since  any image processing operation will introduce artifacts into the whole image rather than the local region within an image. Thus, it is expected that TanH should be better than ReLu, which fits our experimental results very well.

\subsection{Robustness Analysis}
In this subsection, two other experiments have been considered, including evaluating the performances of the proposed model for smaller images and evaluating the pre-trained model on different data source. The details are as follows.

\subsubsection{Evaluation on Different Image Sizes}

\begin{table}[!t]
\renewcommand{\arraystretch}{1.0}
\caption{The average classification accuracies(\%) for different image sizes. The best accuracy among the five methods is highlighted and labeled with an asterisk (*).}
\label{table_different_size}
\centering
\begin{tabular}{ccccc}
\hline
Image size & 256 & 128 & 64 & 32 \\
\hline
The proposed method & \boldblue{98.75*} &  \boldblue{96.48*} & \boldblue{91.33*} & \boldblue{81.46*} \\
Li's method \cite{li2016identification} & 96.31 & 92.39 & 86.08 & 75.40 \\
Chen's method \cite{chen2015median} & 91.53 &  85.48 & 75.24 & 58.75 \\
Bayar's method \cite{bayar2016deep} & 87.22 & 80.34 & 69.31 & 48.75 \\
Xu's method \cite{xu2016structural} & 85.92 & 80.01 & 69.54 & 57.07 \\

\hline

\end{tabular}
\end{table}

In this experiment, we first crop the center part of the images used previously with three different sizes, including  $128\times128$, $64\times64$ and $32\times32$ respectively. The classification accuracies for different images sizes are shown in Table \ref{table_different_size}. From Table \ref{table_different_size}, we can observe that the performance of the five methods drops with decreasing the image size due to insufficient statistics. However, the proposed model always outperforms the others, and the improvement seems larger when the image size is small. For instance, when the image size is $32\times32$, the proposed method still obtains over 81\% average detection results, while the Li' method becomes 75.40\% and the three other CNN based methods become less than 59\%.

\subsubsection{Evaluation on Different Image Sources}

\begin{table}[!t]
\renewcommand{\arraystretch}{1.0}
\caption{The average classification accuracies(\%) for different image sizes on the BOSSbase.}
\label{table_different_size_bossbase}
\centering
\begin{tabular}{ccccc}
\hline
Image size & 256 & 128 & 64 & 32 \\
\hline
The proposed method & 98.14 & 95.77 & 90.34 & 80.58 \\
\hline

\end{tabular}
\end{table}

In this experiment, we  evaluate the proposed model on other testing image source. To this end, we firstly collect 10,000 images from BOSSbase 1.01 \cite{bas2011break}, and then we used the pre-trained model with the image set as mentioned in \ref{Sec:Experiment}-A to test the images from BOSSbase. The experimental results for multiple classification are shown in the Table \ref{table_different_size_bossbase}. From Table \ref{table_different_size_bossbase}, we can observe that the accuracies drop slightly (less than 1\% for all cases) compared with those results shown in Table \ref{table_different_size}, which means that the generalization of the proposed model for image processing operation classification is very good.

\section{Conclusion}
\label{Sec:Conclusion}
In this paper, we proposed a new CNN based method to identify eleven typical image processing operations. The main contributions of this paper are as follows:

\begin{itemize}

\item Considering the property of the investigated forensic problem, we carefully design the architecture of the CNN based model, and provide extensive comparative results to show that the proposed method can achieve state-of-the-art results.

\item To validate of the proposed model, we analyze the influence of different designs in our model, including the high pass filter bank, channel expansion layer, the last pooling layer and the activation functions. Besides, we provide some experimental results to show the robustness of our model.

\end{itemize}

In the future, we will extend our work to identify other image processing operations, such as anti-forensics operations. Furthermore, the robustness against lossy JPEG compression will be considered.

%\section*{Acknowledgment}

\ifCLASSOPTIONcaptionsoff
  \newpage
\fi

\bibliographystyle{IEEEtran}
\bibliography{cnn_forensics}

\begin{thebibliography}{10}
\providecommand{\url}[1]{#1}
\csname url@samestyle\endcsname
\providecommand{\newblock}{\relax}
\providecommand{\bibinfo}[2]{#2}
\providecommand{\BIBentrySTDinterwordspacing}{\spaceskip=0pt\relax}
\providecommand{\BIBentryALTinterwordstretchfactor}{4}
\providecommand{\BIBentryALTinterwordspacing}{\spaceskip=\fontdimen2\font plus
\BIBentryALTinterwordstretchfactor\fontdimen3\font minus
  \fontdimen4\font\relax}
\providecommand{\BIBforeignlanguage}[2]{{%
\expandafter\ifx\csname l@#1\endcsname\relax
\typeout{** WARNING: IEEEtran.bst: No hyphenation pattern has been}%
\typeout{** loaded for the language `#1'. Using the pattern for}%
\typeout{** the default language instead.}%
\else
\language=\csname l@#1\endcsname
\fi
#2}}
\providecommand{\BIBdecl}{\relax}
\BIBdecl

\bibitem{StammWL2013}
M.~C. Stamm, M.~Wu, and K.~R. Liu, ``Information forensics: An overview of the
  first decade,'' \emph{IEEE Access}, vol.~1, pp. 167--200, 2013.

\bibitem{fan2003identification}
Z.~Fan and R.~L. De~Queiroz, ``Identification of bitmap compression history:
  {JPEG} detection and quantizer estimation,'' \emph{IEEE Transactions on Image
  Processing}, vol.~12, no.~2, pp. 230--235, 2003.

\bibitem{farid2009exposing}
H.~Farid, ``Exposing digital forgeries from {JPEG} ghosts,'' \emph{IEEE
  transactions on information forensics and security}, vol.~4, no.~1, pp.
  154--160, 2009.

\bibitem{luo2010jpeg}
W.~Luo, J.~Huang, and G.~Qiu, ``{JPEG} error analysis and its applications to
  digital image forensics,'' \emph{IEEE Transactions on Information Forensics
  and Security}, vol.~5, no.~3, pp. 480--491, 2010.

\bibitem{bianchi2012detection}
T.~Bianchi and A.~Piva, ``Detection of nonaligned double {JPEG} compression
  based on integer periodicity maps,'' \emph{IEEE Transactions on Information
  Forensics and Security}, vol.~7, no.~2, pp. 842--848, 2012.

\bibitem{stamm2008blind}
M.~Stamm and K.~R. Liu, ``Blind forensics of contrast enhancement in digital
  images,'' in \emph{15th IEEE International Conference on Image
  Processing}.\hskip 1em plus 0.5em minus 0.4em\relax IEEE, 2008, pp.
  3112--3115.

\bibitem{cao2014contrast}
G.~Cao, Y.~Zhao, R.~Ni, and X.~Li, ``Contrast enhancement-based forensics in
  digital images,'' \emph{IEEE transactions on information forensics and
  security}, vol.~9, no.~3, pp. 515--525, 2014.

\bibitem{popescu2005exposing}
A.~C. Popescu and H.~Farid, ``Exposing digital forgeries by detecting traces of
  resampling,'' \emph{IEEE Transactions on signal processing}, vol.~53, no.~2,
  pp. 758--767, 2005.

\bibitem{mahdian2008blind}
B.~Mahdian and S.~Saic, ``Blind authentication using periodic properties of
  interpolation,'' \emph{IEEE Transactions on Information Forensics and
  Security}, vol.~3, no.~3, pp. 529--538, 2008.

\bibitem{li2013moment}
L.~Li, J.~Xue, Z.~Tian, and N.~Zheng, ``Moment feature based forensic detection
  of resampled digital images,'' in \emph{Proceedings of the 21st ACM
  international conference on Multimedia}.\hskip 1em plus 0.5em minus
  0.4em\relax ACM, 2013, pp. 569--572.

\bibitem{kirchner2010detection}
M.~Kirchner and J.~J. Fridrich, ``On detection of median filtering in digital
  images.'' in \emph{Media Forensics and Security}, 2010, p. 754110.

\bibitem{yuan2011blind}
H.-D. Yuan, ``Blind forensics of median filtering in digital images,''
  \emph{IEEE Transactions on Information Forensics and Security}, vol.~6,
  no.~4, pp. 1335--1345, 2011.

\bibitem{kang2012robust}
X.~Kang, M.~C. Stamm, A.~Peng, and K.~R. Liu, ``Robust median filtering
  forensics based on the autoregressive model of median filtered residual,'' in
  \emph{Signal \& Information Processing Association Annual Summit and
  Conference}.\hskip 1em plus 0.5em minus 0.4em\relax IEEE, 2012, pp. 1--9.

\bibitem{chen2013blind}
C.~Chen, J.~Ni, and J.~Huang, ``Blind detection of median filtering in digital
  images: A difference domain based approach,'' \emph{IEEE Transactions on
  Image Processing}, vol.~22, no.~12, pp. 4699--4710, 2013.

\bibitem{shi2007steganalysis}
Y.~Q. Shi, C.~Chen, G.~Xuan, and W.~Su, ``Steganalysis versus splicing
  detection,'' in \emph{International Workshop on Digital Watermarking}.\hskip
  1em plus 0.5em minus 0.4em\relax Springer, 2007, pp. 158--172.

\bibitem{he2012digital}
Z.~He, W.~Lu, W.~Sun, and J.~Huang, ``Digital image splicing detection based on
  markov features in {DCT} and {DWT} domain,'' \emph{Pattern Recognition},
  vol.~45, no.~12, pp. 4292--4299, 2012.

\bibitem{zhao2015passive}
X.~Zhao, S.~Wang, S.~Li, and J.~Li, ``Passive image-splicing detection by a
  {2-D} noncausal markov model,'' \emph{IEEE Transactions on Circuits and
  Systems for Video Technology}, vol.~25, no.~2, pp. 185--199, 2015.

\bibitem{qiu2014universal}
X.~Qiu, H.~Li, W.~Luo, and J.~Huang, ``A universal image forensic strategy
  based on steganalytic model,'' in \emph{Proceedings of the 2nd ACM workshop
  on Information hiding and multimedia security}.\hskip 1em plus 0.5em minus
  0.4em\relax ACM, 2014, pp. 165--170.

\bibitem{pevny2010steganalysis}
T.~Pevny, P.~Bas, and J.~Fridrich, ``Steganalysis by subtractive pixel
  adjacency matrix,'' \emph{IEEE Transactions on information Forensics and
  Security}, vol.~5, no.~2, pp. 215--224, 2010.

\bibitem{fridrich2012rich}
J.~Fridrich and J.~Kodovsky, ``Rich models for steganalysis of digital
  images,'' \emph{IEEE Transactions on Information Forensics and Security},
  vol.~7, no.~3, pp. 868--882, 2012.

\bibitem{shi2012textural}
Y.~Q. Shi, P.~Sutthiwan, and L.~Chen, ``Textural features for steganalysis.''
  in \emph{Information hiding}, vol. 7692.\hskip 1em plus 0.5em minus
  0.4em\relax Springer, 2012, pp. 63--77.

\bibitem{fan2015general}
W.~Fan, K.~Wang, and F.~Cayre, ``General-purpose image forensics using patch
  likelihood under image statistical models,'' in \emph{International Workshop
  on Information Forensics and Security}.\hskip 1em plus 0.5em minus
  0.4em\relax IEEE, 2015, pp. 1--6.

\bibitem{li2016identification}
H.~Li, W.~Luo, X.~Qiu, and J.~Huang, ``Identification of various image
  operations using residual-based features,'' \emph{IEEE Transactions on
  Circuits and Systems for Video Technology}, 2016.

\bibitem{lecun1998gradient}
Y.~LeCun, L.~Bottou, Y.~Bengio, and P.~Haffner, ``Gradient-based learning
  applied to document recognition,'' \emph{Proceedings of the IEEE}, vol.~86,
  no.~11, pp. 2278--2324, 1998.

\bibitem{KrizhSH2012}
A.~Krizhevsky, I.~Sutskever, and G.~E. Hinton, ``Imagenet classification with
  deep convolutional neural networks,'' in \emph{Advances in neural information
  processing systems}, 2012, pp. 1097--1105.

\bibitem{simonyan2014very}
K.~Simonyan and A.~Zisserman, ``Very deep convolutional networks for
  large-scale image recognition,'' in \emph{International Conference on
  Learning Representations}, 2015.

\bibitem{qian2015deep}
Y.~Qian, J.~Dong, W.~Wang, and T.~Tan, ``Deep learning for steganalysis via
  convolutional neural networks.'' \emph{Media Watermarking, Security, and
  Forensics}, vol. 9409, pp. 94\,090J--94\,090J, 2015.

\bibitem{xu2016structural}
G.~Xu, H.-Z. Wu, and Y.-Q. Shi, ``Structural design of convolutional neural
  networks for steganalysis,'' \emph{IEEE Signal Processing Letters}, vol.~23,
  no.~5, pp. 708--712, 2016.

\bibitem{ioffe2015batch}
S.~Ioffe and C.~Szegedy, ``Batch normalization: Accelerating deep network
  training by reducing internal covariate shift,'' in \emph{International
  Conference on Machine Learning}, 2015, pp. 448--456.

\bibitem{xu2016ensemble}
G.~Xu, H.-Z. Wu, and Y.~Q. Shi, ``Ensemble of {CNNs} for steganalysis: an
  empirical study,'' in \emph{Proceedings of the 4th ACM Workshop on
  Information Hiding and Multimedia Security}.\hskip 1em plus 0.5em minus
  0.4em\relax ACM, 2016, pp. 103--107.

\bibitem{ni2017deep}
J.~Ni, J.~Ye, and Y.~Yang, ``Deep learning hierarchical representations for
  image steganalysis,'' \emph{IEEE Transactions on Information Forensics and
  Security}, 2017.

\bibitem{chen2015median}
J.~Chen, X.~Kang, Y.~Liu, and Z.~J. Wang, ``Median filtering forensics based on
  convolutional neural networks,'' \emph{IEEE Signal Processing Letters},
  vol.~22, no.~11, pp. 1849--1853, 2015.

\bibitem{bayar2016deep}
B.~Bayar and M.~C. Stamm, ``A deep learning approach to universal image
  manipulation detection using a new convolutional layer,'' in
  \emph{Proceedings of the 4th ACM Workshop on Information Hiding and
  Multimedia Security}.\hskip 1em plus 0.5em minus 0.4em\relax ACM, 2016, pp.
  5--10.

\bibitem{cozzolino2017recasting}
D.~Cozzolino, G.~Poggi, and L.~Verdoliva, ``Recasting residual-based local
  descriptors as convolutional neural networks: an application to image forgery
  detection,'' in \emph{Proceedings of the 5th ACM Workshop on Information
  Hiding and Multimedia Security}.\hskip 1em plus 0.5em minus 0.4em\relax ACM,
  2017, pp. 159--164.

\bibitem{lin2013network}
M.~Lin, Q.~Chen, and S.~Yan, ``Network in network,'' in \emph{International
  Conference on Learning Representations}, 2014.

\bibitem{abadi2016tensorflow}
M.~Abadi, A.~Agarwal, P.~Barham, E.~Brevdo, Z.~Chen, C.~Citro, G.~S. Corrado,
  A.~Davis, J.~Dean, M.~Devin \emph{et~al.}, ``Tensorflow: Large-scale machine
  learning on heterogeneous distributed systems,'' \emph{arXiv preprint
  arXiv:1603.04467}, 2016.

\bibitem{sutskever2013importance}
I.~Sutskever, J.~Martens, G.~Dahl, and G.~Hinton, ``On the importance of
  initialization and momentum in deep learning,'' in \emph{International
  conference on machine learning}, 2013, pp. 1139--1147.

\bibitem{zhan2017image}
Y.~Zhan, Y.~Chen, Q.~Zhang, and X.~Kang, ``Image forensics based on transfer
  learning and convolutional neural network,'' in \emph{Proceedings of the 5th
  ACM Workshop on Information Hiding and Multimedia Security}.\hskip 1em plus
  0.5em minus 0.4em\relax ACM, 2017, pp. 165--170.

\bibitem{rao2016deep}
Y.~Rao and J.~Ni, ``A deep learning approach to detection of splicing and
  copy-move forgeries in images,'' in \emph{International Workshop on
  Information Forensics and Security}.\hskip 1em plus 0.5em minus 0.4em\relax
  IEEE, 2016, pp. 1--6.

\bibitem{nair2010rectified}
V.~Nair and G.~E. Hinton, ``Rectified linear units improve restricted boltzmann
  machines,'' in \emph{Proceedings of the 27th international conference on
  machine learning}, 2010, pp. 807--814.

\bibitem{glorot2011deep}
X.~Glorot, A.~Bordes, and Y.~Bengio, ``Deep sparse rectifier neural networks,''
  in \emph{Proceedings of the Fourteenth International Conference on Artificial
  Intelligence and Statistics}, 2011, pp. 315--323.

\bibitem{bas2011break}
P.~Bas, T.~Filler, and T.~Pevn{\`y}, ``” break our steganographic system”:
  The ins and outs of organizing boss,'' in \emph{Information Hiding}.\hskip
  1em plus 0.5em minus 0.4em\relax Springer, 2011, pp. 59--70.

\end{thebibliography}

\end{document}